\documentclass[apjl]{emulateapj}

\usepackage{amsmath}
\usepackage{amssymb}
\usepackage{amsfonts}
\usepackage{amstext}
\usepackage{graphicx}
\usepackage{fancybox}
\usepackage[usenames,dvipsnames]{color}
\usepackage{pstricks}
\usepackage{bm}
\usepackage{soul, color}
\usepackage{latexsym}
\usepackage{pifont}
\usepackage{shorttoc}
\usepackage{subfigure}
\usepackage{textcomp} 
\usepackage{float}
\usepackage{enumitem}

\usepackage{bm}
\usepackage{lipsum}
\usepackage{threeparttable}
\usepackage{hyperref}
\hypersetup{urlcolor=cyan}

\def\aap{A\&A}

\def\apjs{ApJS}
\def\apjl{ApJL}

\newcommand{\ha}{\rm H$\alpha$\ }

\newcommand{\lya}{Ly$\alpha$\ }

\newcommand{\cm}{\, {\rm cm}}
\newcommand{\angs}{\, {\rm \AA}}

\newcommand{\no}[1]{}

\newcommand{\myemail}{l.m.ribas@astro.uio.no}
\def\lsim{~\rlap{$<$}{\lower 1.0ex\hbox{$\sim$}}}

\def\gsim{~\rlap{$>$}{\lower 1.0ex\hbox{$\sim$}}}

\shorttitle{SSIM: Extended Halos as a Probe of the Ionizing Escape Fraction during the EoR}
\shortauthors{Mas-Ribas, Hennawi, Dijkstra, Davies, Stern \& Rix}

\begin{document}

\title{Small-scale Intensity Mapping:  Extended halos as a Probe of the\\
Ionizing Escape Fraction and Faint Galaxy Populations during Reionization}

\author{Llu\'is Mas-Ribas\altaffilmark{1}} 
\author{Joseph F. Hennawi\altaffilmark{2,3}} 
\author{Mark Dijkstra\altaffilmark{1}}
\author{Frederick B. Davies\altaffilmark{3}} 
\author{Jonathan Stern\altaffilmark{2}} 
\author{Hans-Walter Rix\altaffilmark{2}} 
\altaffiltext{1}{Institute of Theoretical Astrophysics, University of Oslo,
Postboks 1029, 0315 Oslo, Norway.\\
\url{\myemail}}
\altaffiltext{2}{Max-Planck-Institut f{\"u}r Astronomie, K{\"o}nigstuhl 17, 
D-69117 Heidelberg, Germany}
\altaffiltext{3}{Department of Physics, University of California, Santa 
Barbara, CA 93106, USA}


\begin{abstract}

   We present a new method to quantify the value of the escape fraction of ionizing photons, and   
the existence of ultra-faint galaxies clustered around brighter objects during the epoch of 
cosmic reionization, using the diffuse Ly$\alpha$, continuum and \ha emission observed around 
galaxies at $z\sim6$. We model the surface brightness profiles of the diffuse halos 
considering the fluorescent emission powered by ionizing photons escaping from 
the central galaxies, and the nebular emission from satellite star-forming sources, by extending the 
formalisms developed in \cite{Masribas2016} and \cite{Masribas2017}. The comparison between 
our predicted profiles and \lya observations at $z=5.7$ and $z=6.6$ favors a low ionizing escape fraction,  
$f_{\rm esc}^{\rm ion}\sim5\%$, for galaxies in the range $-19\gtrsim M_{\rm UV} \gtrsim -21.5$. 
However, uncertainties and possible systematics in the observations do not allow for firm conclusions. 
We predict \ha and rest-frame visible continuum observations with JWST, and show that JWST
will be able to detect extended (a few tens of kpc) fluorescent H$\alpha$ emission powered by 
ionizing photons escaping from a bright, $L\gsim 5L^*$, galaxy. Such observations can differentiate 
fluorescent emission from nebular emission by satellite sources. We discuss how observations 
and stacking of several objects may provide unique constraints on the escape fraction for faint 
galaxies and/or the abundance of ultra-faint radiation sources.

\end{abstract}

\section{Introduction}

   A fundamental question in the quest for understanding the Epoch of Reionization (EoR) is the 
amount of ionizing radiation released by primeval galaxies into the intergalactic medium (IGM), 
which is at present dominated by two major unknowns: 

    First, the fraction of ionizing photons escaping 
the interstellar (ISM) and circumgalactic (CGM) media, i.e., the ionizing escape fraction, which is  
impossible to measure directly at $z\gtrsim 4$ as the IGM becomes fully opaque to ionizing radiation.   
Various {\it indirect} alternative approaches, including relations between the gas covering fraction and 
reddening \citep[e.g.,][]{Jones2012,Leethochawalit2016,Reddy2016},  H$\beta$ line equivalent width and 
UV spectral slope \citep{Zackrisson2013,Zackrisson2016}, or the analysis of the \lya spectral line profile 
\citep{Dijkstra2016,Verhamme2016} have been proposed, but the values for the escape fraction 
remain within the fairly broad range $0.3 \gtrsim f_{\rm esc}^{\rm ion} \gtrsim 0.01$. Additionally, 
measurements of the Thomson scattering optical depth to the CMB have placed the `galaxy population 
averaged' ionizing escape fraction value within the range $f_{\rm esc}^{\rm ion}\sim0.1-0.2$ 
\citep[e.g.,][]{Robertson2015,Mitra2016,Sun2016}. 

   Second, the number density of faint (down to $-M_{\rm UV}\sim 12 - 10$) galaxies which are often invoked 
to reach the total ionizing photon budget required for reionization. The existence and nature of these 
objects is still highly uncertain: {\it indirect} constraints on their abundance arise from, e.g., lensing, local dwarf 
galaxies and/or Gamma-Ray Burst (GRB) rates studies 
\citep{Kistler2009,Trenti2010,Robertson2012,Kistler2013,Robertson2013,Bouwens2015,Yue2016,
Weisz2017}. 

   In this paper, we propose that extended emission observed around star-forming galaxies 
can provide valuable new insights into both these two unknowns.

   Diffuse extended \lya emission (\lya halos; LAHs) around star-forming galaxies is practically 
ubiquitous at redshifts $2\lesssim z \lesssim6$ \citep[][although see \citealt{Feldmeier2013,
Jiang2013} for non detections]{Steidel2011,Matsuda2012,Momose2014,Wisotzki2015,
Xue2017}. In addition, observations at redshifts $0\lesssim z \lesssim2.5$ demonstrate the (omni-) 
presence of \ha halos, whose extent is usually smaller than that of LAHs but significantly larger than 
the region observed in UV continuum \citep{Hayes2013,Matthee2016,Sobral2017}. The main 
mechanism responsible for the diffuse halos is still not clear, and it is possible that different mechanisms 
dominate at different distances from the center of the galaxies.

   We showed in \cite{Masribas2016} that close to star-forming galaxies, $r\sim20 - 30$ physical 
kpc, fluorescent radiation powered by ionizing radiation escaping from the ISM may contribute up to $50-60\%$ to 
the total \lya surface brightness observed by \cite{Matsuda2012} at $z=3.1$. These values depend in 
detail on the adopted CGM model and parameters for the central galaxy. At larger distances from the 
central galaxy, the predicted fluorescent surface brightness profiles lie below the observations. 
We therefore assessed the contribution of nebular emission produced `in-situ' in satellite sources clustered 
around the central galaxy in \cite{Masribas2017}. These satellites would be too faint to 
be resolved individually, but their integrated emission may be detectable, analogously to the method of 
`intensity mapping' \citep[see][and references therein]{Fonseca2017}, but here applied to much smaller 
scales. In general, these models matched the \lya profiles observed by \cite{Matsuda2012} 
and \cite{Momose2014} at $z=3.1$ remarkably well for different clustering prescriptions. The UV 
continuum profiles, however, appeared usually a factor $1.5 - 3$ above the data, and we required a 
significant evolution of the \lya rest-frame equivalent width with UV magnitude of the sources to recover 
the observed values. 

   Our previous work ignored scattering, because this is extremely sensitive to the physical 
properties, distribution and morphology of the neutral gas, all of which are difficult - and still impossible - 
to model or simulate from first principles \citep[see][]{McCourt2016}.  Irrespective of these uncertainties, 
we expect scattering to smoothen out the surface brightness profiles close to the center, a few 
tens of pkpc. We do not expect scattering to affect the profiles at larger distances \citep[see, e.g.,][]
{Laursen2007,Steidel2011,DijkstraKramer2012,Lake2015}. 

   Finally, gravitational `cooling radiation' can give rise to extended \lya emission \citep{Haiman2000,
DijkstraLoeb2009,Goerdt2010,FaucherGiguere2010,Rosdahl2012,Lake2015}, although predicting 
the cooling luminosity is highly uncertain \citep[e.g.,][]{Yang2006,Cantalupo2008,FaucherGiguere2010}. 
For the typical halo masses hosting Lyman alpha emitters (LAEs), cooling luminosities appear to fall below 
the observed luminosity in halos \citep{Rosdahl2012}, therefore we expect cooling not to be dominant.

 We demonstrated in \cite{Masribas2017} that comparing the surface brightness profiles of the Ly$\alpha$, 
\ha and continuum emission can disentangle the significance of each mechanism. This is because the 
`size' of the halos is connected to the physical properties of the CGM and the radiative processes affecting 
the Ly$\alpha$, \ha and continuum radiation. In detail, continuum radiation (UV \& VIS) has mostly a stellar 
origin and is not affected by radiative transfer effects (scattering), thus precisely tracing the star-forming 
regions. \ha photons arise as a by-product of the recombination of ionized hydrogen, which mostly occurs 
in the HII regions of the ISM (we refer to this as the `nebular component'), but also in gas in the CGM that 
has been ionized by (ionizing) photons that escaped from the ISM (we refer to this as the `fluorescent 
component'). \lya emission occurs mostly in the same regions as \ha but it is a resonant transition, which 
allows the \lya photons to scatter through neutral hydrogen gas, travelling large distances if they do not 
encounter and are destroyed 
by dust. In addition, cooling radiation from cold gas being accreted onto the galaxy 
can give rise to more extended \lya emission compared to H$\alpha$, even in the absence of scattering.
The required surface brightness levels to detect extended emission at wavelengths other than Ly$\alpha$ 
are challenging (observations of extended H$\alpha$ emission are usually at redshifts where we do not 
have access to Ly$\alpha$). Deeper observations from the ground, in combination with observations by the 
James Webb Space Telescope \citep[JWST;][]{Gardner2006} can provide us with a more complete spectral 
coverage of extended emission.

   In this paper, we apply the formalisms we developed in previous works to redshifts $z\sim6$, 
corresponding to the end of the epoch 
of reionization. We expect the fluorescent effect of the central galaxy to clearly dominate the surface 
brightness profiles close to the center because the contribution of the satellite sources depends linearly 
on the cosmic star formation rate, which at these redshifts is much lower than at $z\sim3$. This is 
important because our fluorescent profiles are sensitive to the properties of the central galaxy and the 
CGM, specifically to the escape fraction of ionizing photons and neutral gas covering factor. At larger 
distances, the nebular signature of the satellite sources, if present, may overcome that of the central 
galaxy, thus providing evidence for their existence and relevance to the reionization process.
In \S~\ref{sec:formalism}, we summarize the formalism for the calculation of the surface brightness 
profiles. In \S~\ref{sec:results}, we present our results for \lya (\S~\ref{sec:lya}), and the observational 
strategy and predicted \ha and continuum surface brightness profiles (\S~\ref{sec:havis}). 
Our findings are discussed and we conclude in \S~\ref{sec:discussion}.  

   We assume a flat $\Lambda$CDM cosmology with values ${\rm \Omega_{\Lambda}=
0.7}$, ${\rm \Omega_{ m}}=0.3$ and ${\rm H_0=68\, km\,s^{-1}\,Mpc^{-1}}$.

\section{Formalism}\label{sec:formalism}

   We decompose the total line emission surface brightness profile at impact parameter $b$ 
from the central galaxy into two components, 
\begin{equation}
SB (b) = SB^{\rm cen} (b) + SB^{\rm sat}(b)~, 
\end{equation} 
where $SB^{\rm cen} (b)$ denotes the fluorescent component (discussed in \S~\ref{sec:sbcen}), and 
$SB^{\rm sat}(b)$ the nebular component of satellite sources (discussed in \S~\ref{sec:sbext}).

\subsection{Surface Brightness from the Central Galaxy:\\  Fluorescent emission}\label{sec:sbcen}

  Following \cite{Masribas2016}, we compute the fluorescent surface brightness due to the central 
galaxy as
\begin{equation}\label{eq:sb}
SB^{\rm cen} (b) = \frac{2}{(1+z)^4}\int_{b}^{\infty}\frac{r\,{\rm d}r}{\sqrt{r^2\,-\,b^2}}\, C\,\dot n_{\rm ion}(r)\,f_c(r)\,f_{\rm esc}^{\rm ion} (r) ~.
\end{equation}
Here, $\dot n_{\rm ion}(r)$ denotes the total rate at which ionizing photons are produced by the central 
galaxy divided by $4\pi r^2$, and $C$ represents the \ha and \lya energy 
emitted per ionizing photon and unit of solid angle (see \S~\ref{sec:sfr}). The term $f_c(r)$ is the radial 
gas covering factor, which quantifies the spatial distribution of neutral gas in the CGM, and 
$f_{\rm esc}^{\rm ion}(r)$ denotes the ionizing photon escape fraction (see \S~\ref{sec:medium}). 
In practice, we set the upper limit of the integral in Eq.~\ref{eq:sb} to 100 physical kpc. This choice is 
somewhat arbitrary and motivated by the virial radius of the dark matter halo hosting the central galaxies. 
We tested that variations of a few tens of kpc around this value do not alter our results. We assume that 
line emission produced by fluorescence escapes with $100\%$ efficiency. Values other than this,  
rescale linearly our central galaxy profiles.

 \begin{figure} 
\includegraphics[width=0.47\textwidth]{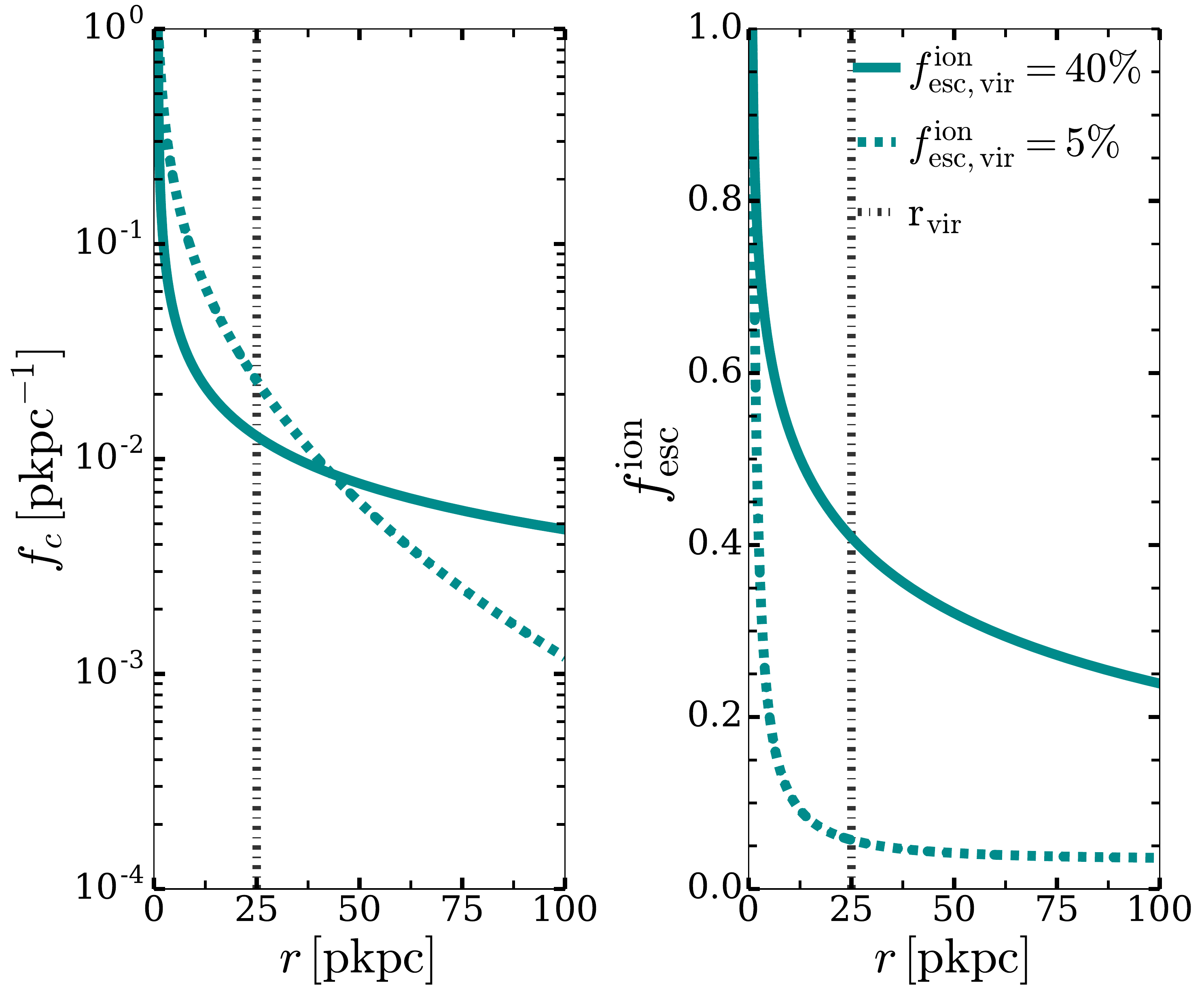}
\caption{{\it Left panel:} Radial profiles of neutral gas covering factor, which denotes 
the number of self-shielding clumps along a differential length at a distance $r$ from the central 
galaxy. {\it Right panel:} Profiles for the escape fraction of ionizing photons, i.e., 
the fraction of the total number of photons reaching a distance $r$ without being absorbed or destroyed. 
The {\it solid and dashed dark cyan lines} represent the two CGM models, parametrized accounting for 
the value of the escape fraction at the virial radius (denoted as {\it dashed vertical line}).} 
\label{fig:fcfesc}
\end{figure}

\subsubsection{Central Galaxy, $C$ and $\dot n_{\rm ion}(r)$}\label{sec:sfr}

   We consider a $5L^*$ galaxy at $z=6.17$ for our calculations, which corresponds to $M_{\rm 
UV}\simeq -22.2$ adopting the fitting formula for $M^*$ from \citealt{Kuhlen2012} (their FIT model). 
We require a bright galaxy for the fluorescent component to dominate above the possible nebular 
satellite signal and avoid contamination, and at a level that can yield a good signal-to-noise ratio (S/N) 
for the observations. As we will show later, targeting fainter galaxies would require the stack of several 
objects  or averaging the signal in larger radial bin sizes in order to obtain reasonable S/N. However, 
brighter galaxies are rare, and the probability of finding one in the field of view (FOV) considered in our 
calculations becomes undesirably small (see \S~\ref{sec:jwst}). The conversion factor between UV 
luminosity and star formation rate (SFR) for high redshift by \cite{Madau2014},
\begin{equation}\label{eq:madau}
L_{\rm UV}{\rm [erg\,s^{-1}\,Hz^{-1}]}  = 8.7\times 10^{27}\,{\rm SFR}\,{\rm [M_\odot \, yr^{-1}]}~,
\end{equation}
yields a SFR $\sim 41.4\,{\rm M_\odot \, yr^{-1}}$ for the central galaxy. We have not considered  
additional dust attenuation for this calculation because we expect the effect of dust at these redshifts to 
be low \citep[see section 3.1.3 in][for further discussion]{Madau2014}. This assumption results in 
a lower limit for the SFR and, in turn, for the surface brightness which depends linearly on SFR. 

   The SFR then determines the total production rate of ionizing photons as \citep{Robertson2013}, 
\begin{equation}\label{eq:lumlya}
\dot N_{\rm ion}\,{\rm [photons\,s^{-1}]} =1.38\times10^{53}\, {\rm SFR\,[M_{\odot}\,yr^{-1}] }~, 
\end{equation}
which we use to compute $n_{\rm ion}(r)$ as
\begin{equation}\label{eq:lumlya}
\dot n_{\rm ion}(r) =  \frac{\dot N_{\rm ion}}{4\pi r^2}~. 
\end{equation}

   We assume that each ionizing photon is converted into H$\alpha$ and Ly$\alpha$ with an efficiency 
that is set by case-B recombination. Under this assumption, the parameter $C$ for \lya and \ha is
\begin{equation}\label{eq:Cpara}
C\,{\rm [erg\,sr^{-1}]} = \frac{ 1}{4\pi }\,\frac{\alpha_{\rm X}^{\rm eff}}{\alpha_{\rm B}}\,h\nu_X~.
\end{equation}
The fraction $\alpha_{\rm X}^{\rm eff}/\alpha_{\rm B}$ denotes the number of line photons per ionizing 
photon 
(X takes on `H$\alpha$' or `Ly$\alpha$'), and equals $0.68$ for Ly$\alpha$ and $0.45$ for H$\alpha$.
In detail, both the production rate of ionizing photons and the conversion efficiency into \ha and \lya line 
photons depend on the IMF, metallicity, stellar populations, etc., at these redshifts \citep[][see also the 
discussion in \citealt{Masribas2017}]{Raiter2010,Masribas2016b}. The factors $h\nu_X$ and $1/4\pi$ 
account for the energy and isotropic emission of the line photons, respectively. 

\subsubsection{CGM, $f_c(r)$ and $f_{\rm esc}^{\rm ion}(r)$}\label{sec:medium}

   The characteristics of the CGM at the redshifts of reionization are not known. We 
parametrize our ignorance by considering the two most extreme CGM models at $z=3.1$ 
developed in \cite{Masribas2016}. In these models, the CGM is composed of a spherically symmetric 
distribution of neutral hydrogen clumps embedded within a hot medium. The distribution of clumps is 
characterized by the cold gas covering fraction, $f_c(r)$, which denotes the number of neutral clumps along 
a differential length, at a distance $r$ from the central galaxy. The escape fraction of ionizing photons 
is distance dependent in our models, and relates to the covering fraction as $f_{\rm esc}^{\rm ion}(r)= {\rm 
exp}[-\int_0^r f_c(r) {\rm d}r]$. To stress, this parameter represents the escape fraction of ionizing photons 
out to a distance $r$. Our calculations assume a $100\%$ escape fraction at $r=0$, but our results 
scale linearly with this value. We parametrize the two CGM 
models using the value of the escape fraction of ionizing photons at the virial radius of the central galaxy 
($\sim 25$ pkpc, for a dark matter halo $M_h \sim1.5\times10^{11}\,{\rm M_\odot}$), and describe them 
below:

\begin{enumerate}[leftmargin=*]

\item $f_{\rm esc,vir}^{\rm ion}=40\%$: This CGM model derives from that proposed by \cite{Steidel2010}, 
where the neutral hydrogen clumps are in pressure equilibrium with a radially accelerating 
outflowing hot medium. The {\it dark cyan solid lines} in the left and right panels of Figure \ref{fig:fcfesc} 
display the gas covering factor and ionizing escape fraction profiles, respectively. This model  
produces an escape fraction at the virial radius of $40\%$. 

\item $f_{\rm esc,vir}^{\rm ion}=5\%$: This CGM model is obtained after applying an inverse 
Abelian transformation to the two-dimensional neutral hydrogen covering factor from the simulations by 
\cite{Rahmati2015}, in order to obtain a radial dependent covering factor (see \citealt{Masribas2016} for 
this transformation). The corresponding profiles are 
represented by the {\it dashed dark cyan lines} in the left and right panels of Figure \ref{fig:fcfesc}.  
The escape fraction in this model reaches a value $\sim5\%$ at the virial radius.

\end{enumerate}

   Figure \ref{fig:fcfesc} shows that the radial dependence of the escape fraction is mostly driven by the 
covering factor profile in the first $\sim50$ pkpc from the central galaxy. The left panel shows that  
the covering factor in the $f_{\rm esc,vir}^{\rm ion}=40\%$  model is smaller at those distances, 
which yields a smoother decrease of the ionizing escape fraction in this case. Conversely, the escape 
fraction profile for the $f_{\rm esc,vir}^{\rm ion}=5\%$ model reaches a low value rapidly at 
$r\sim 25$ pkpc since the covering fraction decreases slowly with radial distance.

\subsection{Surface Brightness from Satellite Sources:\\  Nebular radiation}\label{sec:sbext}

   We calculate the \lya and \ha surface brightness of the satellite sources as in \cite{Masribas2017}, 
\begin{equation} 
SB^{\rm sat}(b)=\frac{2}{(1+z)^4} \int_b^{\infty} \epsilon_X [1+\xi_X (r)] f_{\rm esc,X}\, \frac{r{\rm 
d}r}{\sqrt{r^2-b^2}} ~,
\end{equation} 
where $\epsilon_X$ refers to the average cosmic emissivity for \lya or \ha (see \S~\ref{sec:emiss}), 
$\xi_X$ is the correlation function for the corresponding radiation (see \S~\ref{sec:clust}), and 
$f_{\rm esc,X}$ denotes the escape fraction from the satellites for the transition ${\rm X}$. We (arbitrarily) 
set a fiducial value $f_{\rm esc,X}= 40\%$, but also explore the broad ranges $0.2 \ge f_{\rm esc,X} \ge 
0.7$ and $0.1 \ge f_{\rm esc,X} \ge 1.0$, because the escape fraction is not known at these high redshifts, 
especially for faint satellite sources, and is linked to the uncertain presence of dust,  which 
likely affects more the \lya photons than those of \ha due to radiative transfer effects and 
the frequency dependence of the attenuation curve. We again limit the 
integral to 100 pkpc, and enable the presence of satellites at distances above 10 pkpc. 

   In addition to nebular \lya and \ha radiation, the satellite sources will also produce continuum radiation 
associated to star formation, which will result in an overall extended continuum profile. 
We calculate the visible (rest-frame, VIS) continuum surface brightness profile from that of H$\alpha$ as 
\begin{equation}
SB_{\rm VIS}(b)= \frac{(1+z)}{\rm EW_{H\alpha}}\frac{\lambda_{\rm H{\alpha}}^2}{c}\, SB_{\rm H\alpha}^{\rm sat}(b)   ~.
\end{equation}
We assume a fiducial \ha line equivalent width (rest-frame) ${\rm EW_{H\alpha}=300\,\angs}$,  
considering the observations by \citealt{Queralto2016} (and references therein), and also explore the 
ranges $450\geq {\rm EW_{H\alpha}\,(\AA) \geq 150}$ and $700\geq {\rm EW_{H\alpha}\,(\AA) 
\geq 50}$. 
The parameters $\lambda_{\rm H\alpha}$ and $c$ denote the rest-frame \ha wavelength and the speed of light, 
respectively and, combined with the term $(1+z)$, yield a surface brightness per unit frequency.

\subsubsection{Emissivity, $\epsilon$}\label{sec:emiss}

   To obtain the average \ha cosmic emissivity for the satellite population, we calculate the cosmic SFR 
density at the redshift of interest using equation 2 in \cite{Robertson2015}, which considers radiation from 
sources down to $M_{\rm UV}\sim-13$, and assume that the SFR for the satellites equals this value. 
In reality, the SFR for the satellites will be a fraction of total cosmic star formation, but its value is unknown 
and degenerate with the escape fraction of \ha photons \citep[e.g.,][]{Trenti2012}. We finally use the relation by \cite{Kennicutt2012}, 
\begin{equation}\label{eq:sfrL}
L_{\rm H\alpha}\,{\rm [erg\,s^{-1}]} =1.86\times10^{41}\, {\rm SFR\,[M_{\odot}\,yr^{-1}] }~, 
\end{equation}
to obtain the average volumetric emissivity for \ha radiation. We consider that 
the emissivity for \lya will simply be a factor $8.1$ higher, which is the proper intrinsic ratio 
between the volume emissivities for \lya and \ha computed from the tables in \cite{Osterbrock1989}, 
consistent with the approach adopted in Eq.~\ref{eq:Cpara}, and noting again the 
dependence of this value on dust content. 

\subsubsection{Clustering, $\xi(r)$}\label{sec:clust}

   We adopt a power-law two-point correlation function for the clustering of satellite sources 
around the bright central galaxy, with correlation length $r_0 = 3.79\, {\rm Mpc\,h^{-1}}$ 
and power-law index $\alpha=-1.8$, as reported by \cite{Harikane2016}, for their sample of 
LBG galaxies with average magnitude $\langle M_{\rm UV} \rangle=-19.3$ at $z=5.9$. In 
\cite{Masribas2017} we used another clustering prescription because, at small distances, 
the departures from a power-law by the data of \cite{Ouchi2010} at $z=3.1$ were significant. 
In the current case, however, the power-law approach is consistent within 1$\sigma$ with the 
data by \cite{Harikane2016}, which extend up to $\sim12$ pkpc from the central galaxy. We  
consider the same bias for the \ha radiation and the satellite sources because radiative transfer effects, 
i.e., scattering, are not present for the case of \ha \citep{Masribas2017}. We adopt these same values 
for the satellite clustering when comparing with 
the LAE data by \cite{Momose2014} and \cite{Jiang2013}, owing to the broad range of uncertainties 
and overlap of values between the correlation lengths from $z=5.7$ to  $z=6.8$ in the different data 
samples of LAEs and LBGs in \cite{Ouchi2010} and \cite{Harikane2016}, respectively. 
Possible variations of the clustering of sources and/or radiation are engulfed in the large range explored 
for the escape fraction described above.

\section{Results}\label{sec:results}

   We present below our results. In \S~\ref{sec:lya}, we apply and compare our analytical models to the 
observational \lya data by \cite{Momose2014} and \cite{Jiang2013} at $z=5.7$ and $z=6.6$. In 
\S~\ref{sec:havis}, we detail a potential JWST observational strategy (\S~\ref{sec:jwst}), and present the 
predicted \ha and VIS continuum surface brightness profiles at $z=6.17$ (\S~\ref{sec:predict}). 

\subsection{{\rm Ly$\alpha$}}\label{sec:lya}

\begin{figure*}
\includegraphics[width=0.49\textwidth]{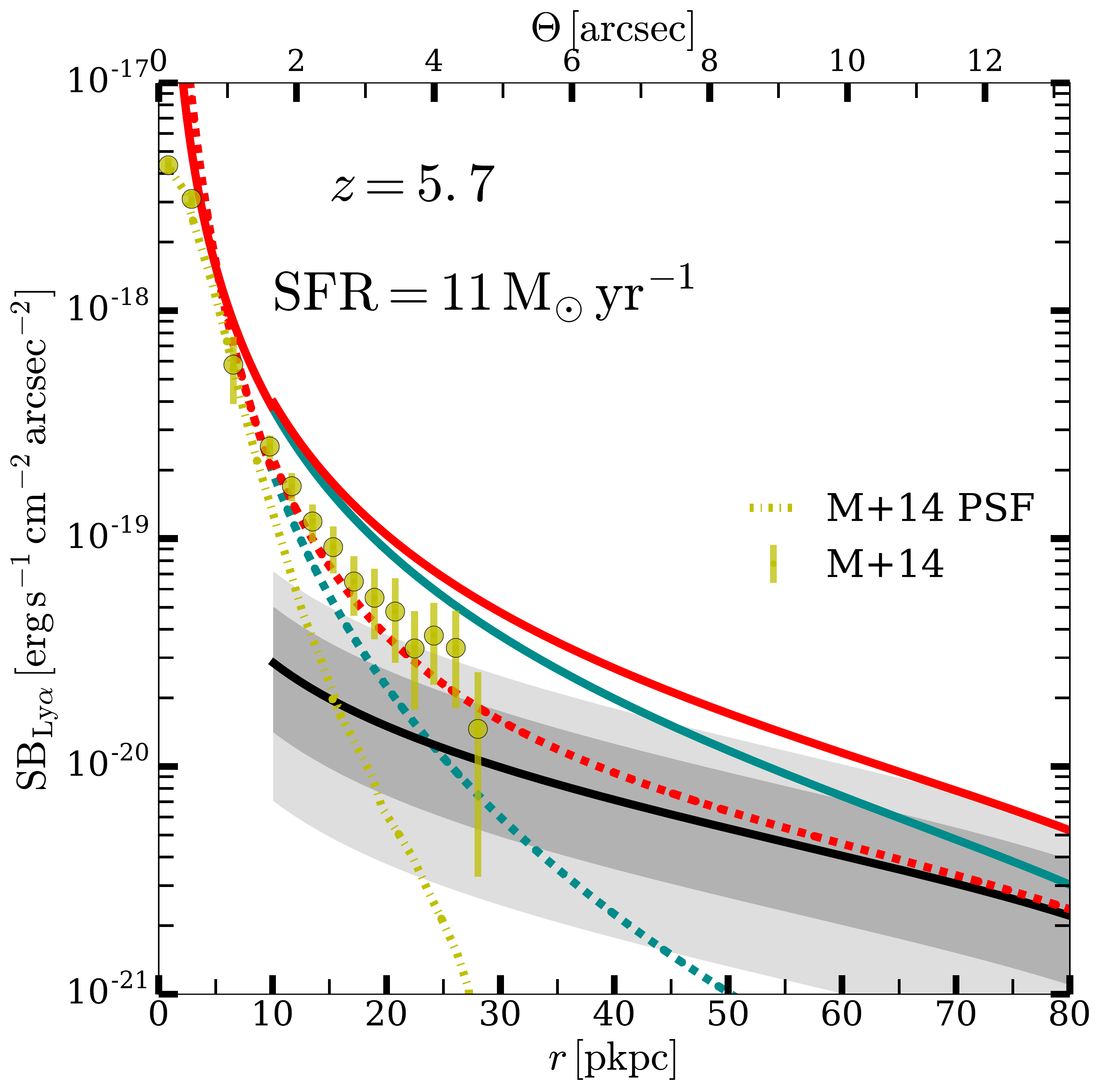}\includegraphics[width=0.464\textwidth]{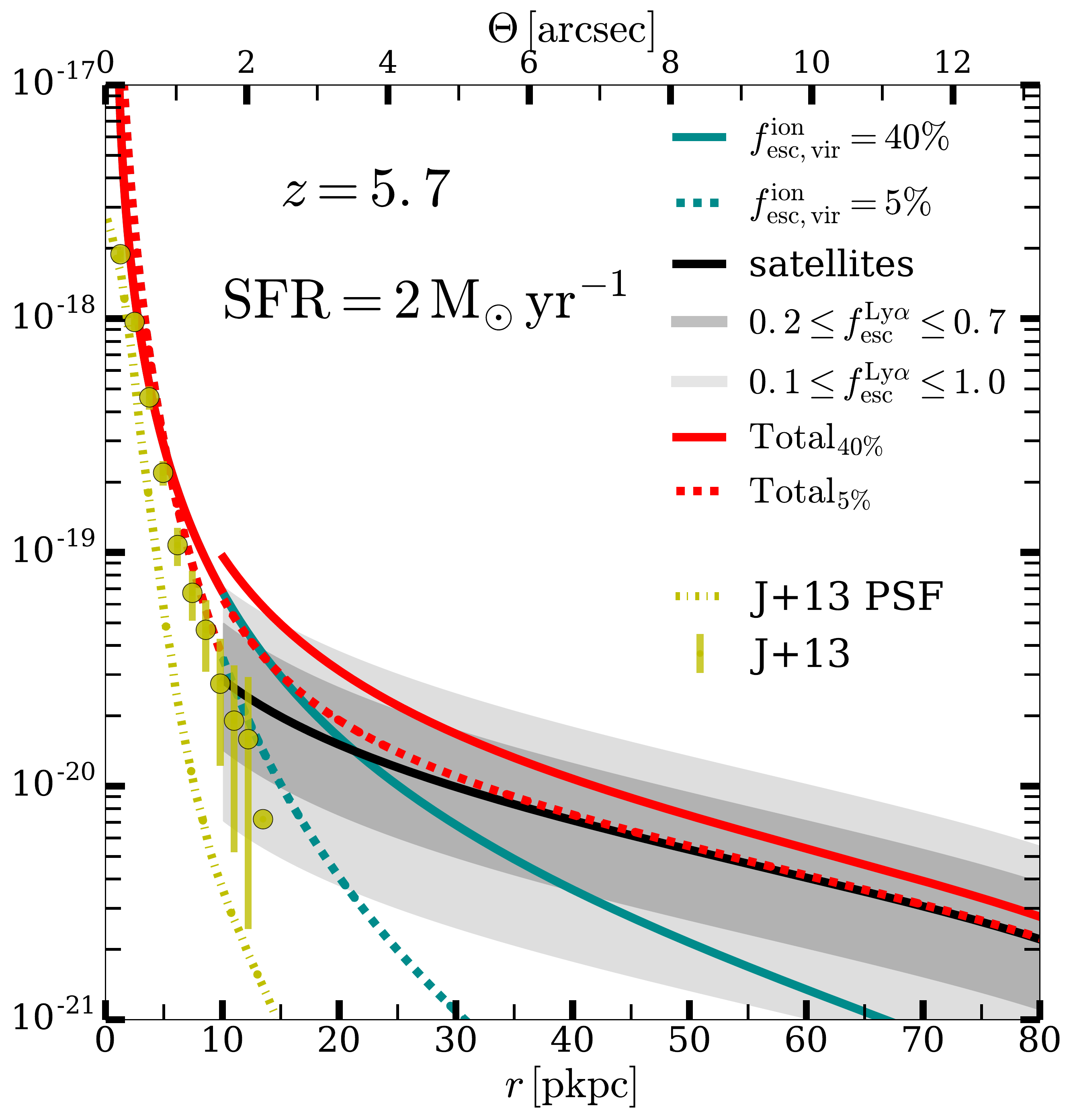}
\caption{Comparison between our predicted \lya surface brightness profiles and the observational data 
by \citealt{Momose2014} (their mean values, M+14, {\it left panel}) and \citealt{Jiang2013} (J+13, {\it 
right panel}) at 
$z=5.7$. The {\it yellow points with uncertainties} denote the observed profiles and the {\it yellow dashed 
lines} the PSFs. The profiles of the $f_{\rm esc,vir}^{\rm ion}=40\%$ ($f_{\rm esc,vir}^{\rm ion}=5\%$)  
CGM models are displayed as the {\it solid} ({\it dashed}) {\it dark cyan lines}. The {\it solid black line} and 
{\it shaded grey bands} represent the fiducial model for the profiles of the satellites and the different 
escape fraction ranges, respectively. The {\it solid (dashed) red lines} denote the total (central + satellite) 
profiles for the $40\%$ ($5\%$) CGM models. The SFRs for the central galaxies are quoted in every panel. 
In general, the low escape fraction models result in a better match to both data sets, although the 
contribution of satellites is unclear (see text).}
\label{fig:57}
\end{figure*}

\begin{figure*}
\includegraphics[width=0.49\textwidth]{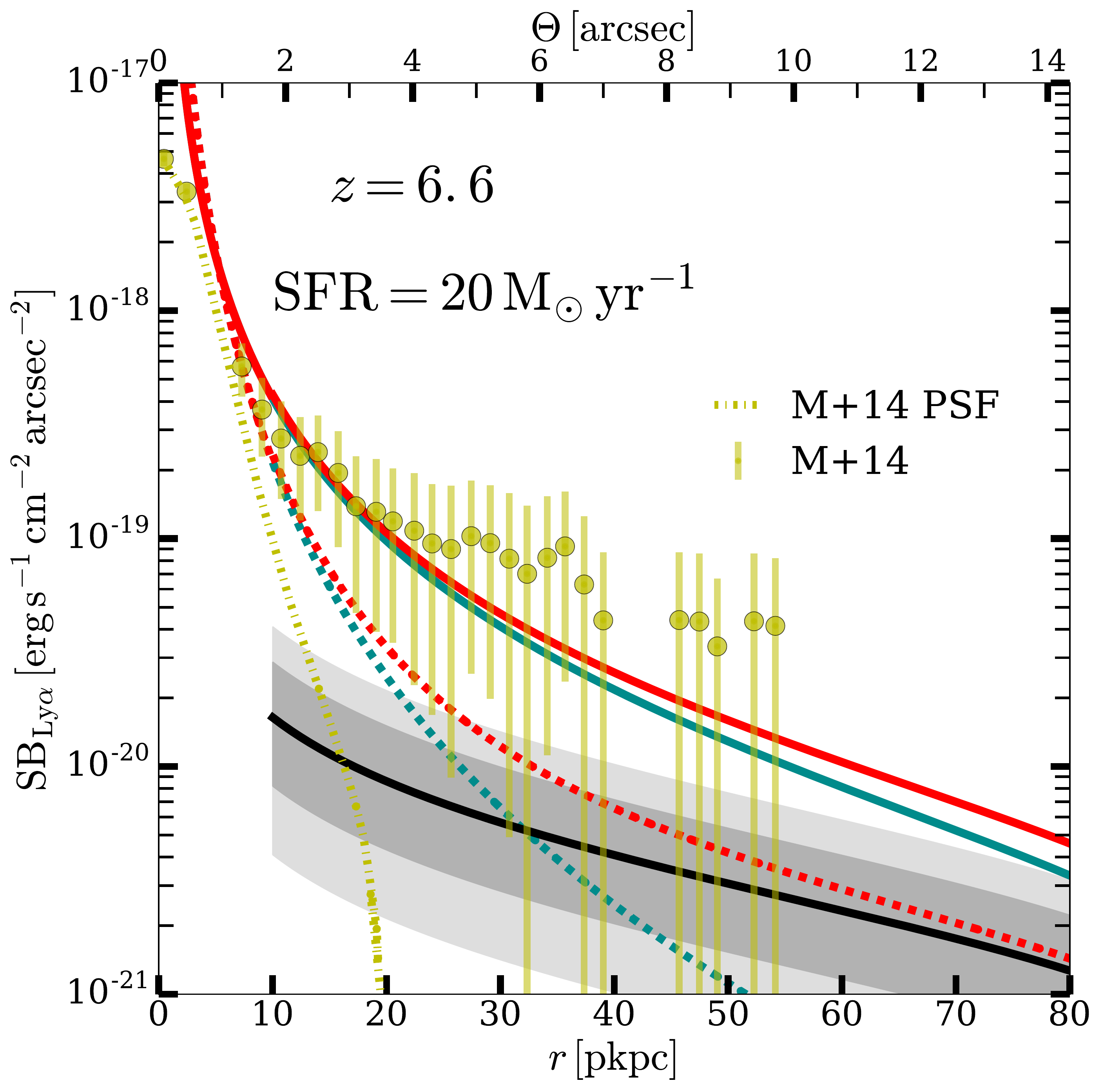}\includegraphics[width=0.464\textwidth]{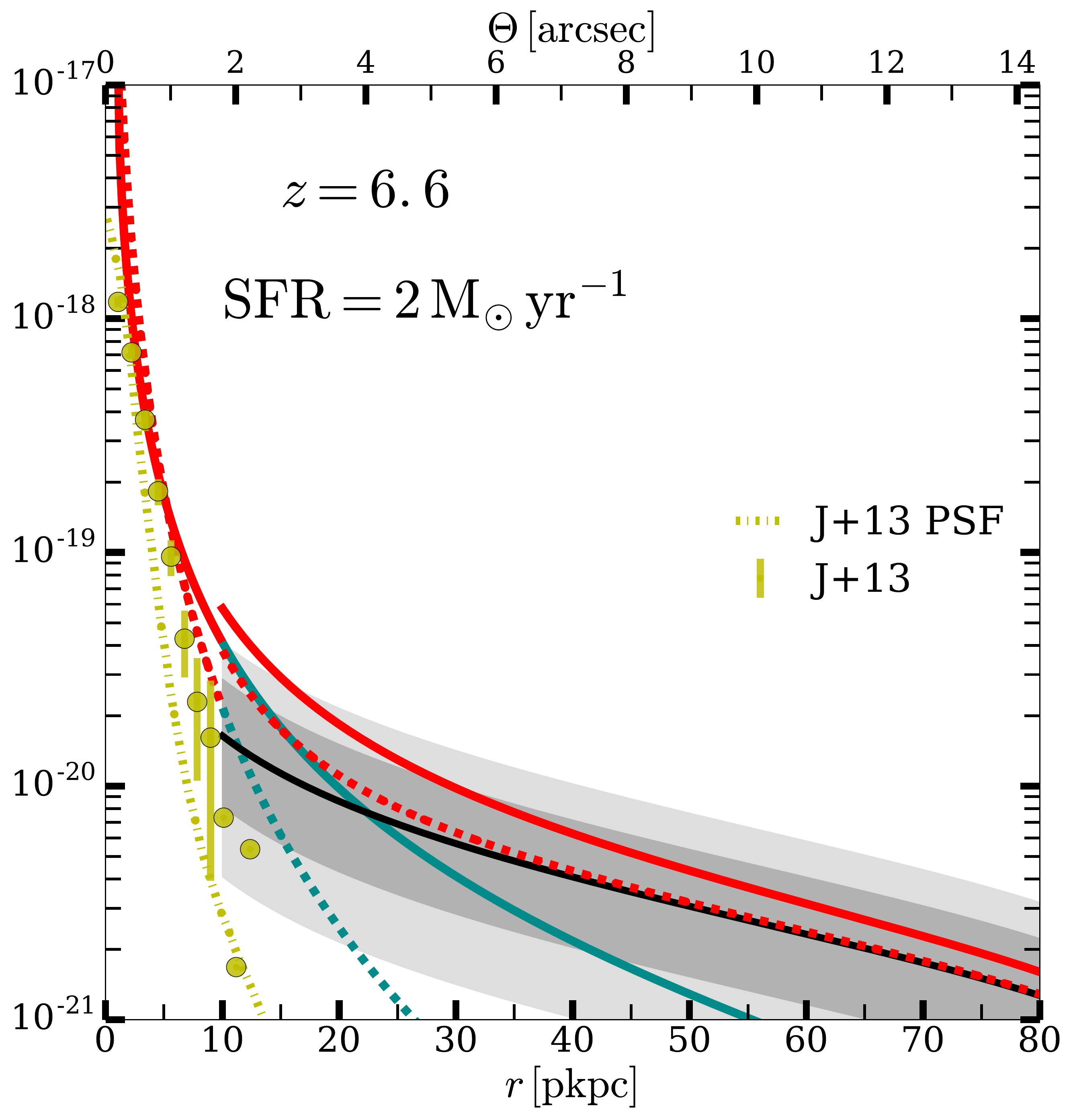}
\caption{Same as Figure \ref{fig:57} but at $z=6.6$. In this case, the data by Momose et al. (2014) seems 
to be better matched by the model with high ionizing escape fraction, although the error bars are consistent 
with the two CGM models within 1$\sigma$ for most of the data points, and the possibility for the presence 
of artifacts in the observations is elevated (see text). As at $z=5.7$, the data by Jiang et al. (2013) favors 
the low escape fraction model and discards the contribution from satellites. }
\label{fig:66}
\end{figure*}

   Figures \ref{fig:57} and \ref{fig:66} display the comparison between our predicted \lya surface brightness 
profiles and the observational data at $z=5.7$ and $z=6.6$, respectively, by Momose et al. 2014 (their 
mean values, {\it left panels}) and Jiang et al. 2013 ({\it right panels}). The {\it yellow data points and error 
bars} denote the data and uncertainties, and the {\it yellow dashed lines} the point-spread functions 
(PSFs) quoted by these authors. In general, the data by Momose et al. show significant extended 
emission well beyond the PSF. Jiang et al. argue that the extended emission in their observations is 
due to the fact that the central galaxy is resolved, but not to the presence 
of an extended halo. The \lya data by Jiang et al. corresponds to the stack of 43 (40) LAEs at $z=5.7$ 
($z=6.6$), a subsample selected from the observations by \cite{Ouchi2008,Ouchi2010,Kashikawa2011}, 
reaching a surface brightness limit $\sim1.2\times10^{-19}\,{\rm erg\,s^{-1}\cm^{-2}\,arcsec^{-2}}$. The 
stacked profiles in Momose et al. arise from 397 (119) LAEs at $z=5.7$ ($z=6.6$), selected from the parent 
sample of observations by \cite{Ouchi2008,Ouchi2010}, and reach similar (within a factor $\sim2$) \lya 
surface brightness limits compared to the Jiang et al. data. The two datasets of overlapping (but 
different) galaxy samples do not agree with each other; Momose et al. attributes the lack of extended 
emission in the analysis by Jiang et al. to the small sample size.

   To obtain consistent comparisons, we calculate the star formation rates for the central galaxy in 
the observed data sets as follows: We integrate the corresponding UV continuum surface brightness profiles by 
\citealt{Momose2014} (their figure 3) to calculate the UV luminosity and then use Eq.~\ref{eq:madau} to 
obtain the SFRs. Since the UV surface brightness profiles or galaxy sample are not available in 
\cite{Jiang2013}, we calculate the ratio between the integrated \lya surface brightness profiles in 
the inner regions, where the central galaxy dominates, in \cite{Momose2014} and 
\cite{Jiang2013}, and assume these \lya ratios to be the same as for the SFRs. At $z=5.7$, integrating 
up to a distance of 4 (1) arcsec, the ratio is $\sim5.5$ ($\sim4$). At $z=6.6$, the same upper limits 
for the integral result in ratios $\sim12.8$ and $\sim 6.6$, respectively. For simplicity, we set 
the ratios to 5.5 and 10 at $z=5.7$ and $z=6.6$, respectively, and stress that the results depend linearly 
on these parameters. 

   The {\it dark cyan solid (dashed) lines} in Figures \ref{fig:57} and \ref{fig:66} denote the surface 
brightness profiles for the $f_{\rm esc,vir}^{\rm ion}=40\%$ ($f_{\rm esc,vir}^{\rm ion}=5\%$) CGM models,  
with the SFR of the central galaxy quoted in each panel. The {\it black lines and shaded grey areas} 
represent the profiles for the fiducial model and the two ranges of nebular radiation escape fraction, 
respectively, for the satellite sources. The {\it red solid (dashed) lines} represent the total, central + 
satellite galaxies, surface brightness profiles for the $40\%$ ($5\%$) CGM models. 

   At $z=5.7$ (Figure \ref{fig:57}), both data sets are better matched by the low escape fraction CGM 
model, despite the factor $5.5$ between the respective SFRs. The contribution of satellites appears, 
however, unclear: For the Jiang et al. data (right panel), the match to the observations 
is the best when satellites are ignored while, for the Momose et al. case (left panel), the satellites 
provide the necessary signal to reach the observed values at distances $r\gtrsim15$ pkpc. 

   At $z=6.6$ (Figure \ref{fig:66}), the data by Jiang et al. (right panel) is better reproduced by the low 
escape fraction CGM model again but, contrary to the previous findings, for the Momose et al. data (left 
panel), even the high escape fraction model lies below the observations in this case, and the contribution 
of the satellite sources is not enough to approach the observed levels. However, the uncertainties in the 
data by Momose et al. here are significantly large, resulting in the low escape fraction model being 
at a $\lesssim1\sigma$ level for most of the data points. The high observed profile might be a consequence 
of the high star formation rate, ${\rm SFR\sim20\,M_{\odot}\,yr^{-1}}$, but it is more likely that the 
observations below $\sim1-2\times10^{-19}\,{\rm erg\,s^{-1}\cm^{-2}\,arcsec^{-2}}$ are mostly driven  
by systematic effects, as tested by \citealt{Momose2014} (their section 3.2 and figure 8). 
 
   Taking into account the possible systematic effects in the data by \cite{Momose2014} at $z=6.6$, the low 
ionizing escape fraction CGM model, $f_{\rm esc,vir}^{\rm ion}=5\%$, seems to be the most consistent one 
with the observed \lya profiles in general, regardless the average SFR values for the central galaxies, which 
cover the corresponding magnitude range $-19\gtrsim M_{\rm UV} \gtrsim -21.5$. If this result is confirmed, 
these galaxies would not have contributed significantly to the reionization process, and a large 
population of fainter galaxies with a high ionizing escape fraction would be necessary. However, 
the presence of satellite sources is never required by the data in \cite{Jiang2013}, and only at distances 
$r\gtrsim15-20$ physical kpc from the central galaxy by the observations of \cite{Momose2014}, where 
the uncertainties and possible systematic effects are important. The apparent discrepancy between the 
two different data-sets prevents us from drawing firm conclusions. We present below the H$\alpha$ 
and visual continuum predicted profiles and JWST observations, which will constitute additional probes, 
and will further extend this discussion in \S~\ref{sec:discussion}.

\subsection{\ha and VIS}\label{sec:havis}

   We present below the predicted \ha and visible (VIS) continuum surface brightness profiles 
(\S~\ref{sec:predict}), and the approach we adopt for future JWST observations of these extended 
halos (\S~\ref{sec:jwst}). Contrary to Ly$\alpha$, the \ha transition is not resonant, implying that the \ha 
photons do not scatter the neutral hydrogen gas. This characteristic facilitates the interpretation of  
the \ha profiles compared to those of Ly$\alpha$: \ha photons, either come from star formation or 
fluorescence, thus tracing their production sites.

\subsubsection{\ha \& VIS: JWST Observational Strategy}\label{sec:jwst}

   We consider the near infrared camera, NIRCam, instrument onboard JWST for our observations, 
with a field of view ${\rm FOV=9.68}\, {\rm arcmin^2}$. We perform the calculations at $z=6.17$ for 
the \ha radiation to match the position of the F470N narrow-band (NB) filter, with a band width (BW) 
${\rm BW=0.051\,\mu}$m, and centered at $4.708\,\mu$m observer-frame. For the visible continuum, we 
adopt the filter F410M, with ${\rm BW=0.438\,\mu}$m, and centered at $4.082\,\mu$m, corresponding to 
a rest-frame wavelength $\sim 5693\,{\rm \AA}$. In both cases, we consider a total observing time 
$t_{\rm exp}=10^5$ s. We compute the sky background at $4.708\,\mu$m following the technical note 
\url{http://www.stsci.edu/~tumlinso/ nrs_sens_2852.pdf} (Eq.~22), resulting in a surface brightness 
$SB_{\rm sky}=6\,\mu {\rm Jy\, arcsec^{-2}\,(\sim 22\,mag\,arcsec^{-2})}$. 

\begin{figure*}
\includegraphics[width=0.47\textwidth]{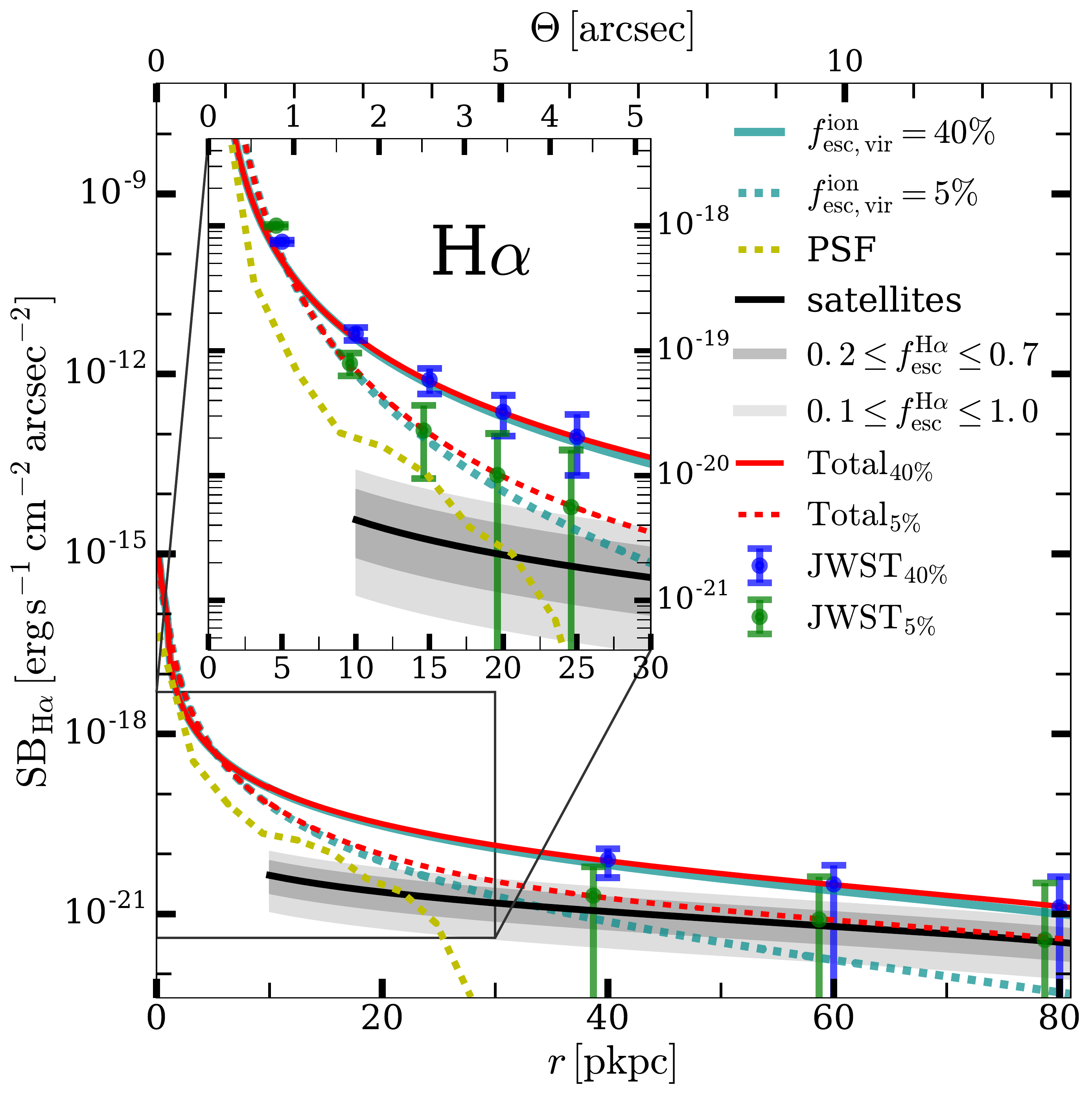}\includegraphics[width=0.50\textwidth]{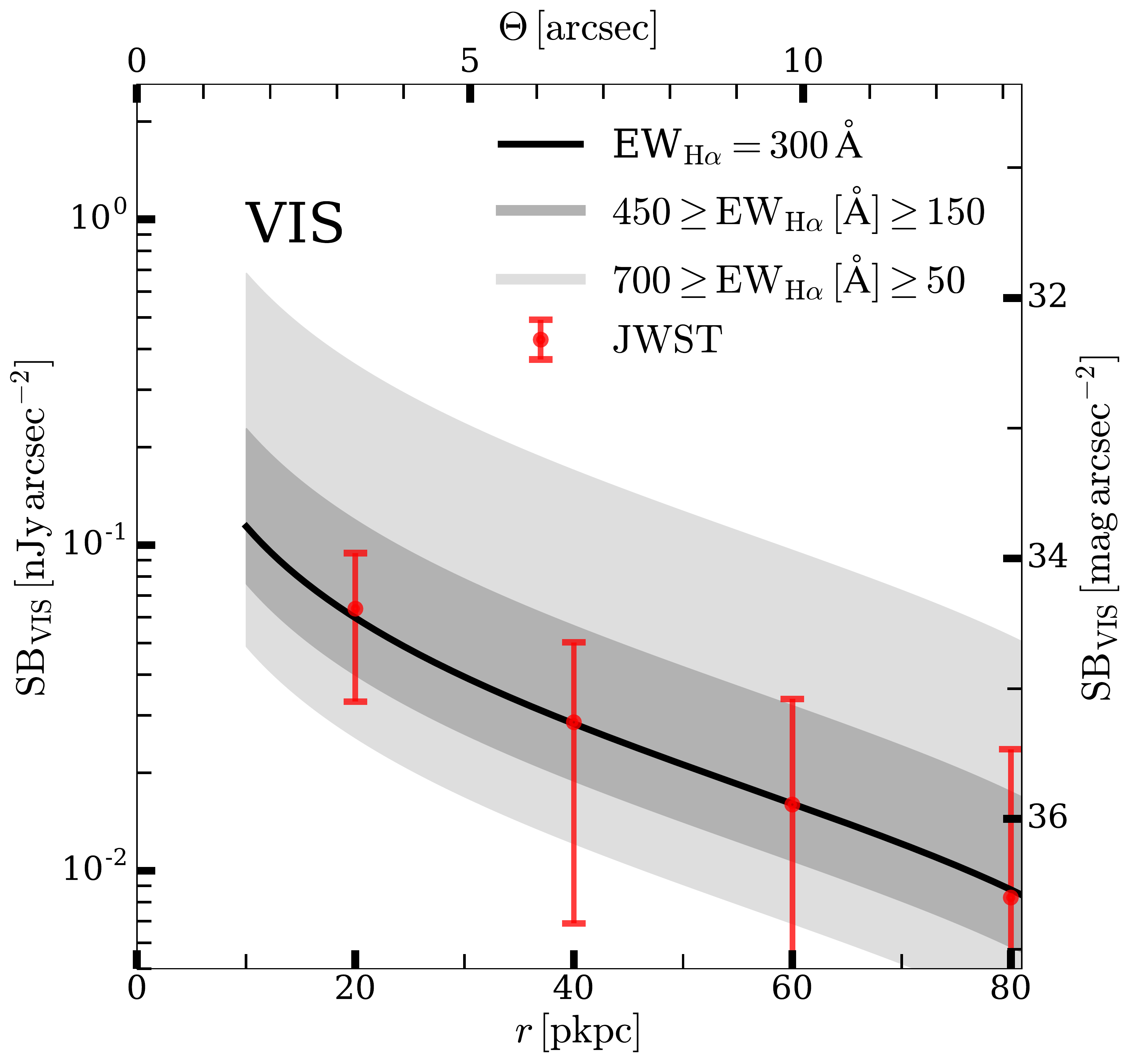}
\caption{ {\it Left panel:} Radial surface brightness profiles with physical distance for one $5L^*$ galaxy. 
The {\it solid} ({\it dashed}) {\it dark cyan lines} denote the \ha profiles of the 
$f_{\rm esc,vir}^{\rm ion}=40\%$ ($f_{\rm esc,vir}^{\rm ion}=5\%$) CGM models. The {\it black line} 
represents the profile from the satellite sources considering $f_{\rm esc}^{\rm H\alpha}=40\%$, while the 
{\it grey shaded areas} denote the ranges $0.2 \ge f_{\rm esc}^{\rm H\alpha} \ge 0.7$ and $0.1 \ge 
f_{\rm esc}^{\rm H\alpha} \ge 1.0$. The {\it solid (dashed) red line} denotes the total (central + satellite) 
profile for the $40\%$ ($5\%$) CGM model, and the {\it blue} and {\it green data points} with their 
uncertainty, are the predicted corresponding JWST observations for these total profiles. The {\it inset 
figure} zooms in the inner $r \le 30$ pkpc. The green vertical lines denoting uncertainties in the 
observations are slightly shifted from the original position to facilitate the visualization. 
{\it Right panel:} Radial visible (VIS) continuum profile due to the satellite sources, obtained after 
stacking the images of $\sim 18\,L>0.1 L^*$ galaxies detected in one NIRCam pointing. 
The {\it black solid line} denotes the fiducial model (${\rm EW_{H\alpha}=300\,\angs}$), 
and the shaded regions the equivalent width ranges $450\geq {\rm EW_{H\alpha}\,(\AA) \geq 150}$ and 
$700\geq {\rm EW_{H\alpha}\,(\AA) \geq 50}$. The observations and uncertainties are displayed in red.} 
\label{fig:sb}
\end{figure*}

   Following \citealt{Masribas2017} (Appendix B), we derive the uncertainties in the observations from 
the signal-to-noise ratio (S/N), computed as ${\rm S/N}=N_s / \sqrt{N_s+N_{\rm sky}}$, where $N_s$ 
and $N_{\rm sky}$ are the azimuthally integrated source and sky photon number counts, respectively, 
defined as
\begin{eqnarray}
N_s &=& \frac{f_{\rm H\alpha}}{h\,\nu_{\rm H\alpha}^{\rm obs}}\, {\rm A_{ aper}}\, \eta\, t_{\rm exp} ~,\\
N_{\rm sky} &=& \frac{f_{\rm sky}}{h\,\nu_{\rm H\alpha}^{\rm obs}}\,{\rm BW}\, {\rm A_{ aper}}\, \eta\, t_{\rm exp} ~.
\end{eqnarray}
The aperture for JWST is ${\rm A_{ aper}=25\,m^2}$, and $\eta=0.469\, (0.274)$ is the total system 
throughput for the continuum (line) filter. The terms ${f_{\rm H\alpha}}$ and ${f_{\rm sky}}$ are the 
source and sky fluxes, 
respectively, resulting from the integration of the surface brightness over the area of the corresponding 
radial annulus around the central galaxy, and ${h\nu_{\rm H\alpha}^{\rm obs}}$ denotes the \ha photon 
energy at the observer frame, $z=6.17$. We include the term BW in the sky background 
and VIS calculations because the sky and continuum are in units of flux density. We have not 
accounted for other systematics or noise effects in this simple calculation but we have checked that 
our S/N results are in broad agreement with those produced using the JWST on-line Time Exposure 
Calculator\footnote{\url{https://jwst.etc.stsci.edu/}} (ETC).

   We obtain the profile driven by the PSF as follows: We compute the encircled energy (EE) radial profile 
for the narrow band F470N filter with the publicly available package 
\texttt{WebbPSF}\footnote{\url{http://pythonhosted.org/webbpsf/}}, and convolve it with the total \ha 
luminosity produced by the central galaxy, assumed to be a point source. We plot the resulting PSF 
profile as the {\it yellow dashed line} in Figure \ref{fig:sb}, after applying the luminosity distance and 
geometric factors to transform the luminosity into surface brightness.

   To obtain a high S/N for the \ha profiles, it would be desirable to observe the brightest possible galaxies, 
$L \gsim 10 L^*$, but these objects are sufficiently rare that it is unlikely to find one in a single JWST NIRCam 
pointing (observation). Fainter objects, $L\sim L^*$, result in larger number densities but, in this case, the signal 
from possible satellites easily overcomes that of the central galaxy, contaminating the escape fraction 
measurements. In addition, the stacking of a large number of these objects would be required to reach a good 
S/N, introducing possible systematics from the complex stacking methodology, and increasing the number 
of JWST pointings, i.e., observing time. We consider the case of observing one $5L^*$ galaxy previously 
detected by a wide-field spectroscopic or NB imaging survey covering an area in the sky of 10 deg$^2$.   
We adopt these numbers as a compromise between the number of $5L^*$ objects in the volume adopted, 
$\sim 2$ when integrating the UV luminosity function at $z=6.17$ with the parameters from the fitting formula 
by \citealt{Kuhlen2012} ($M_{\rm UV}^*=-20.374$, $\phi^*=9.5\times10^{\rm -4}\,{\rm Mpc^{-3}\, mag^{-1}}$, 
and $\alpha=-1.85$), the required observing time, and the S/N. 

    The signal from the satellite sources is  
independent of the brightness of the central galaxy, assuming that the clustering is similar for a range 
of galaxy luminosities. We consider the stacked image of $\sim 18\,L>0.1 L^*\,(M_{\rm {UV}}<-18)$ 
galaxies for the calculation of the visible continuum observation, which is the number of objects present in 
the NIRCam FOV (a single pointing).

\subsubsection{\ha \& VIS: Predicted Surface Brightness Profiles}\label{sec:predict}

   We present the resulting \ha and continuum surface brightness profiles, considering one 
$5L^*$ and eighteen $L>0.1L^*$ galaxies, respectively, at $z=6.17$. The analysis of the continuum 
will be useful to probe the satellite contribution for both, assessing their existence and separating 
their contribution from that of the central galaxy. 
The {\it left panel} in Figure \ref{fig:sb} displays the surface brightness profiles for H$\alpha$, 
where the {\it dark cyan solid (dashed) line} denotes the predictions for the $f_{\rm esc,vir}^{\rm ion}=
40\%$ ($f_{\rm esc,vir}^{\rm ion}=5\%$) CGM model for the central galaxy. 
The {\it black line} represents the fiducial profile for the satellite sources at $r\ge10$ pkpc, considering 
$f_{\rm esc}^{\rm H\alpha}=40\%$, and the {\it grey shaded areas} represent 
the ranges $0.2 \ge f_{\rm esc}^{\rm H\alpha} \ge 0.7$ and $0.1 \ge f_{\rm esc}^{\rm H\alpha} 
\ge 1.0$. The {\it solid (dashed) red line} denotes the total (central + satellites)   
profile for the $40\%$ ($5\%$) CGM model, and the {\it blue} and {\it green error bars} represent the 
uncertainties in the JWST observations described in the previous section for the corresponding total 
profiles. The {\it yellow dashed line} denotes the PSF profile. The {\it inset figure} 
zooms into the range $5 \lesssim r \lesssim 30$ pkpc. The {\it right panel} shows 
the surface brightness profiles and observations of the visible continuum from the satellite galaxies, for 
the fiducial model (${\rm EW_{H\alpha}=300\,\angs}$), and the equivalent width ranges $450\geq 
{\rm EW_{H\alpha}\,(\AA) \geq 150}$ and $700\geq {\rm EW_{H\alpha}\,(\AA) \geq 50}$. The {\it red data 
and error bars} denote the JWST observations, which present detections up to $\sim 40$ pkpc from the 
center of the galaxy.

   The left panel in Figure \ref{fig:sb} indicates that JWST observations will be able to probe the 
surface brightness profiles up to distances $r\sim 50$ pkpc for the case of the $f_{\rm esc}^{\rm 
ion}\sim40\%$ model. These observations would imply that a significant fraction of the ionizing radiation 
from galaxies reaches large distances from the center, contributing to the reionization of the 
intergalactic medium (IGM). Instead, a low ionizing escape fraction similar to the $f_{\rm esc}^{\rm ion}
\sim5\%$ model, results in a profile detectable up to $r\sim 15$ pkpc. This result would indicate that 
most of the ionizing radiation does not escape beyond the CGM, thus 
not contributing significantly to cosmic reionization.  

   The fluorescent profile of the central galaxies, especially for the low ionizing escape fraction cases, 
might be contaminated by the PSF and the nebular radiation from the satellite sources. Observations of the 
VIS continuum profile will contribute differentiating between the two processes. A steep (almost non-existent) 
continuum profile beyond the central galaxy would imply that satellite radiation (and sources) are 
not important, and the extended profiles are driven by fluorescence. Conversely, an extended continuum 
signal would imply the presence of star formation (satellite sources) in the halo of the central galaxy, 
and would allow for the investigation of the nature and contribution of these faint source population to 
reionization. It is likely that the faint sources are too faint to be individually detected but their collective 
emission would produce the observable predicted profile, resembling the intensity mapping methodology.
The right panel in Figure \ref{fig:sb}, however, demonstrates that the observation and stacking of 
several sources is necessary to obtain high S/N and differentiate between the different escape fraction 
scenarios for the satellite sources. Alternatively, larger radial bins sizes than the ones adopted here 
would provide higher S/N values at the cost of spatial resolution.

\section{Discussion and Conclusion}\label{sec:discussion}

   We have proposed that extended emission around galaxies at $z\sim6$ can be used to constrain the 
escape fraction of the ionizing radiation and/or the presence of faint satellite galaxies 
during the late stages of cosmic reionization. We have predicted radial surface brightness profiles 
which include the contribution from ({\it i}) fluorescent emission powered by ionizing radiation leaking from the 
central galaxy \citep[as in][for two different CGM prescriptions]{Masribas2016}, and ({\it ii}), nebular emission 
from faint satellite sources that possibly resided within the halo of the central galaxies \citep[as in][]{Masribas2017}. 
We have compared our predictions with observations of \lya halos at $z=5.7$ 
and $z=6.6$, and have also predicted \ha and visible continuum surface brightness profiles which may be 
detectable within future JWST observations.

   We present and discuss our findings below:
   
\begin{itemize}[leftmargin=*]

\item Our comparison of models to observations favors the model with a very low escape fraction of ionizing  
radiation, $f_{\rm esc}^{\rm ion}\sim 5\%$, from galaxies within the range of magnitudes $-19\gtrsim M_{\rm 
UV} \gtrsim -21.5$. This scenario implies that these galaxies do not contribute significantly to the reionization 
process. Therefore, if galaxies are the major contributors to the reionization process, as recently 
restated by \cite{Parsa2017}, a large population of faint sources with a high ionizing escape fraction 
is necessary. 

\item The presence of faint satellite sources is unclear. 
Two out of four comparisons with current \lya data indicate that satellite sources might 
contribute to the extended \lya emission, but the 
effect of systematics and uncertainties in the observations do not allow for a clear conclusion. Given the 
required existence of faint sources driven by the low escape fraction results, if the presence of satellite 
sources around brighter galaxies is ruled out by the observations, this might imply that the population 
of faint objects is spread out within the intergalactic medium, making its detection more challenging. 

\item JWST will be able to probe the ionizing escape fraction from one bright, 
$L\gsim5L^*$, galaxy up to distances of a few tens of pkpc from the center. Fainter galaxies 
and/or a high signal-to-noise ratio for the continuum observations require the observation 
and stacking of several objects.

\end{itemize}

   We stress that we have adopted and extrapolated two CGM models that were derived to match 
observables at redshifts around $z\sim3$. The validity of our results, therefore, depends on the ability of 
these two models to realistically describe the medium surrounding $z\sim6$ galaxies. While a more detailed 
parametrization of the CGM at high redshift (if currently feasible) is beyond the scope of our work, future 
observations of extended emission together with absorption studies \citep[e.g.,][see also 
\citealt{Steidel2010,Steidel2011}]{DijkstraKramer2012,HennawiProchaska2013,Prochaska2013} will clearly enable 
us to use our proposed method to further constrain the escape fraction.  

   The prospects for doing this experiment at lower redshifts, $z \sim 3-4$, are interesting as the 
surface brightness for emission lines depends on redshift as $(1+z)^4$. However, the contribution from 
satellites can be more important at these low redshifts and overlap with the signal from the central 
galaxy \citep{Masribas2017}. At $z \sim3.5$, we may be also benefiting from the observations by 
HETDEX \citep{Hill2008} or MUSE \citep{Bacon2014}, for constraints on \lya halos. 
While the escape fraction of ionizing photons at $z\sim3.5$ is not directly relevant for 
reionization, using our approach may provide new insights for constraining $f_{\rm esc}^{\rm ion}$ and, 
in turn, the physical properties of the circumgalactic medium.

\section*{acknowledgements}
LMR is grateful to the ENIGMA group at the MPIA in Heidelberg for kind hospitality and 
inspiring discussions. Thanks to Chuck Steidel and Charlotte A. Mason for useful discussions 
about the use of \ha radiation for high-z studies, Erick Zackrisson for clarifying several 
observational aspects and Dan Weisz for an inspiring discussion on the presence of 
faint galaxies during reionization. LMR and MD are grateful to the Astronomy Department at  
UCSB for kind hospitality.

\bibliographystyle{apj}
\bibliography{Ha_halo}\label{References}

\begin{thebibliography}{}
\expandafter\ifx\csname natexlab\endcsname\relax\def\natexlab#1{#1}\fi

\bibitem[{{Bacon} {et~al.}(2014){Bacon}, {Vernet}, {Borisova}, {Bouch{\'e}},
  {Brinchmann}, {Carollo}, {Carton}, {Caruana}, {Cerda}, {Contini}, {Franx},
  {Girard}, {Guerou}, {Haddad}, {Hau}, {Herenz}, {Herrera}, {Husemann},
  {Husser}, {Jarno}, {Kamann}, {Krajnovic}, {Lilly}, {Mainieri}, {Martinsson},
  {Palsa}, {Patricio}, {P{\'e}contal}, {Pello}, {Piqueras}, {Richard},
  {Sandin}, {Schroetter}, {Selman}, {Shirazi}, {Smette}, {Soto}, {Streicher},
  {Urrutia}, {Weilbacher}, {Wisotzki}, \& {Zins}}]{Bacon2014}
{Bacon}, R., {Vernet}, J., {Borisova}, E., {et~al.} 2014, The Messenger, 157,
  13

\bibitem[{{Bouwens} {et~al.}(2015){Bouwens}, {Illingworth}, {Oesch}, {Trenti},
  {Labb{\'e}}, {Bradley}, {Carollo}, {van Dokkum}, {Gonzalez}, {Holwerda},
  {Franx}, {Spitler}, {Smit}, \& {Magee}}]{Bouwens2015}
{Bouwens}, R.~J., {Illingworth}, G.~D., {Oesch}, P.~A., {et~al.} 2015, \apj,
  803, 34

\bibitem[{{Cantalupo} {et~al.}(2008){Cantalupo}, {Porciani}, \&
  {Lilly}}]{Cantalupo2008}
{Cantalupo}, S., {Porciani}, C., \& {Lilly}, S.~J. 2008, \apj, 672, 48

\bibitem[{{Dijkstra} {et~al.}(2016){Dijkstra}, {Gronke}, \&
  {Venkatesan}}]{Dijkstra2016}
{Dijkstra}, M., {Gronke}, M., \& {Venkatesan}, A. 2016, \apj, 828, 71

\bibitem[{{Dijkstra} \& {Kramer}(2012)}]{DijkstraKramer2012}
{Dijkstra}, M., \& {Kramer}, R. 2012, \mnras, 424, 1672

\bibitem[{{Dijkstra} \& {Loeb}(2009)}]{DijkstraLoeb2009}
{Dijkstra}, M., \& {Loeb}, A. 2009, \mnras, 400, 1109

\bibitem[{{Faucher-Gigu{\`e}re} {et~al.}(2010){Faucher-Gigu{\`e}re}, {Kere{\v
  s}}, {Dijkstra}, {Hernquist}, \& {Zaldarriaga}}]{FaucherGiguere2010}
{Faucher-Gigu{\`e}re}, C.-A., {Kere{\v s}}, D., {Dijkstra}, M., {Hernquist},
  L., \& {Zaldarriaga}, M. 2010, \apj, 725, 633

\bibitem[{{Feldmeier} {et~al.}(2013){Feldmeier}, {Hagen}, {Ciardullo},
  {Gronwall}, {Gawiser}, {Guaita}, {Hagen}, {Bond}, {Acquaviva}, {Blanc},
  {Orsi}, \& {Kurczynski}}]{Feldmeier2013}
{Feldmeier}, J.~J., {Hagen}, A., {Ciardullo}, R., {et~al.} 2013, \apj, 776, 75

\bibitem[{{Fonseca} {et~al.}(2017){Fonseca}, {Silva}, {Santos}, \&
  {Cooray}}]{Fonseca2017}
{Fonseca}, J., {Silva}, M.~B., {Santos}, M.~G., \& {Cooray}, A. 2017, \mnras,
  464, 1948

\bibitem[{{Gardner} {et~al.}(2006){Gardner}, {Mather}, {Clampin}, {Doyon},
  {Greenhouse}, {Hammel}, {Hutchings}, {Jakobsen}, {Lilly}, {Long}, {Lunine},
  {McCaughrean}, {Mountain}, {Nella}, {Rieke}, {Rieke}, {Rix}, {Smith},
  {Sonneborn}, {Stiavelli}, {Stockman}, {Windhorst}, \& {Wright}}]{Gardner2006}
{Gardner}, J.~P., {Mather}, J.~C., {Clampin}, M., {et~al.} 2006, \ssr, 123, 485

\bibitem[{{Goerdt} {et~al.}(2010){Goerdt}, {Dekel}, {Sternberg}, {Ceverino},
  {Teyssier}, \& {Primack}}]{Goerdt2010}
{Goerdt}, T., {Dekel}, A., {Sternberg}, A., {et~al.} 2010, \mnras, 407, 613

\bibitem[{{Haiman} {et~al.}(2000){Haiman}, {Spaans}, \&
  {Quataert}}]{Haiman2000}
{Haiman}, Z., {Spaans}, M., \& {Quataert}, E. 2000, \apjl, 537, L5

\bibitem[{{Harikane} {et~al.}(2016){Harikane}, {Ouchi}, {Ono}, {More}, {Saito},
  {Lin}, {Coupon}, {Shimasaku}, {Shibuya}, {Price}, {Lin}, {Hsieh}, {Ishigaki},
  {Komiyama}, {Silverman}, {Takata}, {Tamazawa}, \& {Toshikawa}}]{Harikane2016}
{Harikane}, Y., {Ouchi}, M., {Ono}, Y., {et~al.} 2016, \apj, 821, 123

\bibitem[{{Hayes} {et~al.}(2013){Hayes}, {{\"O}stlin}, {Schaerer}, {Verhamme},
  {Mas-Hesse}, {Adamo}, {Atek}, {Cannon}, {Duval}, {Guaita}, {Herenz}, {Kunth},
  {Laursen}, {Melinder}, {Orlitov{\'a}}, {Ot{\'{\i}}-Floranes}, \&
  {Sandberg}}]{Hayes2013}
{Hayes}, M., {{\"O}stlin}, G., {Schaerer}, D., {et~al.} 2013, \apjl, 765, L27

\bibitem[{{Hennawi} \& {Prochaska}(2013)}]{HennawiProchaska2013}
{Hennawi}, J.~F., \& {Prochaska}, J.~X. 2013, \apj, 766, 58

\bibitem[{{Hill} {et~al.}(2008){Hill}, {Gebhardt}, {Komatsu}, {Drory},
  {MacQueen}, {Adams}, {Blanc}, {Koehler}, {Rafal}, {Roth}, {Kelz}, {Gronwall},
  {Ciardullo}, \& {Schneider}}]{Hill2008}
{Hill}, G.~J., {Gebhardt}, K., {Komatsu}, E., {et~al.} 2008, in Astronomical
  Society of the Pacific Conference Series, Vol. 399, Panoramic Views of Galaxy
  Formation and Evolution, ed. T.~{Kodama}, T.~{Yamada}, \& K.~{Aoki}, 115

\bibitem[{{Jiang} {et~al.}(2013){Jiang}, {Egami}, {Fan}, {Windhorst}, {Cohen},
  {Dav{\'e}}, {Finlator}, {Kashikawa}, {Mechtley}, {Ouchi}, \&
  {Shimasaku}}]{Jiang2013}
{Jiang}, L., {Egami}, E., {Fan}, X., {et~al.} 2013, \apj, 773, 153

\bibitem[{{Jones} {et~al.}(2012){Jones}, {Stark}, \& {Ellis}}]{Jones2012}
{Jones}, T., {Stark}, D.~P., \& {Ellis}, R.~S. 2012, \apj, 751, 51

\bibitem[{{Kashikawa} {et~al.}(2011){Kashikawa}, {Shimasaku}, {Matsuda},
  {Egami}, {Jiang}, {Nagao}, {Ouchi}, {Malkan}, {Hattori}, {Ota}, {Taniguchi},
  {Okamura}, {Ly}, {Iye}, {Furusawa}, {Shioya}, {Shibuya}, {Ishizaki}, \&
  {Toshikawa}}]{Kashikawa2011}
{Kashikawa}, N., {Shimasaku}, K., {Matsuda}, Y., {et~al.} 2011, \apj, 734, 119

\bibitem[{{Kennicutt} \& {Evans}(2012)}]{Kennicutt2012}
{Kennicutt}, R.~C., \& {Evans}, N.~J. 2012, \araa, 50, 531

\bibitem[{{Kistler} {et~al.}(2009){Kistler}, {Y{\"u}ksel}, {Beacom}, {Hopkins},
  \& {Wyithe}}]{Kistler2009}
{Kistler}, M.~D., {Y{\"u}ksel}, H., {Beacom}, J.~F., {Hopkins}, A.~M., \&
  {Wyithe}, J.~S.~B. 2009, \apjl, 705, L104

\bibitem[{{Kistler} {et~al.}(2013){Kistler}, {Yuksel}, \&
  {Hopkins}}]{Kistler2013}
{Kistler}, M.~D., {Yuksel}, H., \& {Hopkins}, A.~M. 2013, ArXiv e-prints,
  arXiv:1305.1630

\bibitem[{{Kuhlen} \& {Faucher-Gigu{\`e}re}(2012)}]{Kuhlen2012}
{Kuhlen}, M., \& {Faucher-Gigu{\`e}re}, C.-A. 2012, \mnras, 423, 862

\bibitem[{{Lake} {et~al.}(2015){Lake}, {Zheng}, {Cen}, {Sadoun}, {Momose}, \&
  {Ouchi}}]{Lake2015}
{Lake}, E., {Zheng}, Z., {Cen}, R., {et~al.} 2015, ArXiv e-prints,
  arXiv:1502.01349

\bibitem[{{Laursen} \& {Sommer-Larsen}(2007)}]{Laursen2007}
{Laursen}, P., \& {Sommer-Larsen}, J. 2007, \apjl, 657, L69

\bibitem[{{Leethochawalit} {et~al.}(2016){Leethochawalit}, {Jones}, {Ellis},
  {Stark}, \& {Zitrin}}]{Leethochawalit2016}
{Leethochawalit}, N., {Jones}, T.~A., {Ellis}, R.~S., {Stark}, D.~P., \&
  {Zitrin}, A. 2016, \apj, 831, 152

\bibitem[{{Madau} \& {Dickinson}(2014)}]{Madau2014}
{Madau}, P., \& {Dickinson}, M. 2014, \araa, 52, 415

\bibitem[{{M{\'a}rmol-Queralt{\'o}} {et~al.}(2016){M{\'a}rmol-Queralt{\'o}},
  {McLure}, {Cullen}, {Dunlop}, {Fontana}, \& {McLeod}}]{Queralto2016}
{M{\'a}rmol-Queralt{\'o}}, E., {McLure}, R.~J., {Cullen}, F., {et~al.} 2016,
  \mnras, 460, 3587

\bibitem[{{Mas-Ribas} \& {Dijkstra}(2016)}]{Masribas2016}
{Mas-Ribas}, L., \& {Dijkstra}, M. 2016, \apj, 822, 84

\bibitem[{{Mas-Ribas} {et~al.}(2016){Mas-Ribas}, {Dijkstra}, \&
  {Forero-Romero}}]{Masribas2016b}
{Mas-Ribas}, L., {Dijkstra}, M., \& {Forero-Romero}, J.~E. 2016, \apj, 833, 65

\bibitem[{{Mas-Ribas} {et~al.}(2017){Mas-Ribas}, {Dijkstra}, {Hennawi},
  {Trenti}, {Momose}, \& {Ouchi}}]{Masribas2017}
{Mas-Ribas}, L., {Dijkstra}, M., {Hennawi}, J.~F., {et~al.} 2017, \apj, 841, 19

\bibitem[{{Matsuda} {et~al.}(2012){Matsuda}, {Yamada}, {Hayashino}, {Yamauchi},
  {Nakamura}, {Morimoto}, {Ouchi}, {Ono}, {Umemura}, \& {Mori}}]{Matsuda2012}
{Matsuda}, Y., {Yamada}, T., {Hayashino}, T., {et~al.} 2012, \mnras, 425, 878

\bibitem[{{Matthee} {et~al.}(2016){Matthee}, {Sobral}, {Oteo}, {Best}, {Smail},
  {R{\"o}ttgering}, \& {Paulino-Afonso}}]{Matthee2016}
{Matthee}, J., {Sobral}, D., {Oteo}, I., {et~al.} 2016, \mnras, 458, 449

\bibitem[{{McCourt} {et~al.}(2016){McCourt}, {Oh}, {O'Leary}, \&
  {Madigan}}]{McCourt2016}
{McCourt}, M., {Oh}, S.~P., {O'Leary}, R.~M., \& {Madigan}, A.-M. 2016, ArXiv
  e-prints, arXiv:1610.01164

\bibitem[{{Mitra} {et~al.}(2016){Mitra}, {Choudhury}, \& {Ferrara}}]{Mitra2016}
{Mitra}, S., {Choudhury}, T.~R., \& {Ferrara}, A. 2016, ArXiv e-prints,
  arXiv:1606.02719

\bibitem[{{Momose} {et~al.}(2014){Momose}, {Ouchi}, {Nakajima}, {Ono},
  {Shibuya}, {Shimasaku}, {Yuma}, {Mori}, \& {Umemura}}]{Momose2014}
{Momose}, R., {Ouchi}, M., {Nakajima}, K., {et~al.} 2014, \mnras, 442, 110

\bibitem[{{Osterbrock}(1989)}]{Osterbrock1989}
{Osterbrock}, D.~E. 1989, {Astrophysics of gaseous nebulae and active galactic
  nuclei}

\bibitem[{{Ouchi} {et~al.}(2008){Ouchi}, {Shimasaku}, {Akiyama}, {Simpson},
  {Saito}, {Ueda}, {Furusawa}, {Sekiguchi}, {Yamada}, {Kodama}, {Kashikawa},
  {Okamura}, {Iye}, {Takata}, {Yoshida}, \& {Yoshida}}]{Ouchi2008}
{Ouchi}, M., {Shimasaku}, K., {Akiyama}, M., {et~al.} 2008, \apjs, 176, 301

\bibitem[{{Ouchi} {et~al.}(2010){Ouchi}, {Shimasaku}, {Furusawa}, {Saito},
  {Yoshida}, {Akiyama}, {Ono}, {Yamada}, {Ota}, {Kashikawa}, {Iye}, {Kodama},
  {Okamura}, {Simpson}, \& {Yoshida}}]{Ouchi2010}
{Ouchi}, M., {Shimasaku}, K., {Furusawa}, H., {et~al.} 2010, \apj, 723, 869

\bibitem[{{Parsa} {et~al.}(2017){Parsa}, {Dunlop}, \& {McLure}}]{Parsa2017}
{Parsa}, S., {Dunlop}, J.~S., \& {McLure}, R.~J. 2017, ArXiv e-prints,
  arXiv:1704.07750

\bibitem[{{Prochaska} {et~al.}(2013){Prochaska}, {Hennawi}, {Lee}, {Cantalupo},
  {Bovy}, {Djorgovski}, {Ellison}, {Lau}, {Martin}, {Myers}, {Rubin}, \&
  {Simcoe}}]{Prochaska2013}
{Prochaska}, J.~X., {Hennawi}, J.~F., {Lee}, K.-G., {et~al.} 2013, \apj, 776,
  136

\bibitem[{{Rahmati} {et~al.}(2015){Rahmati}, {Schaye}, {Bower}, {Crain},
  {Furlong}, {Schaller}, \& {Theuns}}]{Rahmati2015}
{Rahmati}, A., {Schaye}, J., {Bower}, R.~G., {et~al.} 2015, ArXiv e-prints,
  arXiv:1503.05553

\bibitem[{{Raiter} {et~al.}(2010){Raiter}, {Schaerer}, \&
  {Fosbury}}]{Raiter2010}
{Raiter}, A., {Schaerer}, D., \& {Fosbury}, R.~A.~E. 2010, \aap, 523, A64

\bibitem[{{Reddy} {et~al.}(2016){Reddy}, {Steidel}, {Pettini},
  {Bogosavljevi{\'c}}, \& {Shapley}}]{Reddy2016}
{Reddy}, N.~A., {Steidel}, C.~C., {Pettini}, M., {Bogosavljevi{\'c}}, M., \&
  {Shapley}, A.~E. 2016, \apj, 828, 108

\bibitem[{{Robertson} \& {Ellis}(2012)}]{Robertson2012}
{Robertson}, B.~E., \& {Ellis}, R.~S. 2012, \apj, 744, 95

\bibitem[{{Robertson} {et~al.}(2015){Robertson}, {Ellis}, {Furlanetto}, \&
  {Dunlop}}]{Robertson2015}
{Robertson}, B.~E., {Ellis}, R.~S., {Furlanetto}, S.~R., \& {Dunlop}, J.~S.
  2015, \apjl, 802, L19

\bibitem[{{Robertson} {et~al.}(2013){Robertson}, {Furlanetto}, {Schneider},
  {Charlot}, {Ellis}, {Stark}, {McLure}, {Dunlop}, {Koekemoer}, {Schenker},
  {Ouchi}, {Ono}, {Curtis-Lake}, {Rogers}, {Bowler}, \&
  {Cirasuolo}}]{Robertson2013}
{Robertson}, B.~E., {Furlanetto}, S.~R., {Schneider}, E., {et~al.} 2013, \apj,
  768, 71

\bibitem[{{Rosdahl} \& {Blaizot}(2012)}]{Rosdahl2012}
{Rosdahl}, J., \& {Blaizot}, J. 2012, \mnras, 423, 344

\bibitem[{{Sobral} {et~al.}(2017){Sobral}, {Matthee}, {Best}, {Stroe},
  {R{\"o}ttgering}, {Oteo}, {Smail}, {Morabito}, \&
  {Paulino-Afonso}}]{Sobral2017}
{Sobral}, D., {Matthee}, J., {Best}, P., {et~al.} 2017, \mnras, 466, 1242

\bibitem[{{Steidel} {et~al.}(2011){Steidel}, {Bogosavljevi{\'c}}, {Shapley},
  {Kollmeier}, {Reddy}, {Erb}, \& {Pettini}}]{Steidel2011}
{Steidel}, C.~C., {Bogosavljevi{\'c}}, M., {Shapley}, A.~E., {et~al.} 2011,
  \apj, 736, 160

\bibitem[{{Steidel} {et~al.}(2010){Steidel}, {Erb}, {Shapley}, {Pettini},
  {Reddy}, {Bogosavljevi{\'c}}, {Rudie}, \& {Rakic}}]{Steidel2010}
{Steidel}, C.~C., {Erb}, D.~K., {Shapley}, A.~E., {et~al.} 2010, \apj, 717, 289

\bibitem[{{Sun} \& {Furlanetto}(2016)}]{Sun2016}
{Sun}, G., \& {Furlanetto}, S.~R. 2016, \mnras, 460, 417

\bibitem[{{Trenti} {et~al.}(2012){Trenti}, {Perna}, {Levesque}, {Shull}, \&
  {Stocke}}]{Trenti2012}
{Trenti}, M., {Perna}, R., {Levesque}, E.~M., {Shull}, J.~M., \& {Stocke},
  J.~T. 2012, \apjl, 749, L38

\bibitem[{{Trenti} {et~al.}(2010){Trenti}, {Stiavelli}, {Bouwens}, {Oesch},
  {Shull}, {Illingworth}, {Bradley}, \& {Carollo}}]{Trenti2010}
{Trenti}, M., {Stiavelli}, M., {Bouwens}, R.~J., {et~al.} 2010, \apjl, 714,
  L202

\bibitem[{{Verhamme} {et~al.}(2016){Verhamme}, {Orlitova}, {Schaerer},
  {Izotov}, {Worseck}, {Thuan}, \& {Guseva}}]{Verhamme2016}
{Verhamme}, A., {Orlitova}, I., {Schaerer}, D., {et~al.} 2016, ArXiv e-prints,
  arXiv:1609.03477

\bibitem[{{Weisz} \& {Boylan-Kolchin}(2017)}]{Weisz2017}
{Weisz}, D.~R., \& {Boylan-Kolchin}, M. 2017, ArXiv e-prints, arXiv:1702.06129

\bibitem[{{Wisotzki} {et~al.}(2016){Wisotzki}, {Bacon}, {Blaizot},
  {Brinchmann}, {Herenz}, {Schaye}, {Bouch{\'e}}, {Cantalupo}, {Contini},
  {Carollo}, {Caruana}, {Courbot}, {Emsellem}, {Kamann}, {Kerutt}, {Leclercq},
  {Lilly}, {Patr{\'{\i}}cio}, {Sandin}, {Steinmetz}, {Straka}, {Urrutia},
  {Verhamme}, {Weilbacher}, \& {Wendt}}]{Wisotzki2015}
{Wisotzki}, L., {Bacon}, R., {Blaizot}, J., {et~al.} 2016, \aap, 587, A98

\bibitem[{{Xue} {et~al.}(2017){Xue}, {Lee}, {Dey}, {Reddy}, {Hong}, {Prescott},
  {Inami}, {Jannuzi}, \& {Gonzalez}}]{Xue2017}
{Xue}, R., {Lee}, K.-S., {Dey}, A., {et~al.} 2017, \apj, 837, 172

\bibitem[{{Yang} {et~al.}(2006){Yang}, {Zabludoff}, {Dav{\'e}}, {Eisenstein},
  {Pinto}, {Katz}, {Weinberg}, \& {Barton}}]{Yang2006}
{Yang}, Y., {Zabludoff}, A.~I., {Dav{\'e}}, R., {et~al.} 2006, \apj, 640, 539

\bibitem[{{Yue} {et~al.}(2016){Yue}, {Ferrara}, \& {Xu}}]{Yue2016}
{Yue}, B., {Ferrara}, A., \& {Xu}, Y. 2016, \mnras, 463, 1968

\bibitem[{{Zackrisson} {et~al.}(2013){Zackrisson}, {Inoue}, \&
  {Jensen}}]{Zackrisson2013}
{Zackrisson}, E., {Inoue}, A.~K., \& {Jensen}, H. 2013, \apj, 777, 39

\bibitem[{{Zackrisson} {et~al.}(2016){Zackrisson}, {Binggeli}, {Finlator},
  {Gnedin}, {Paardekooper}, {Shimizu}, {Inoue}, {Jensen}, {Micheva},
  {Khochfar}, \& {Dalla Vecchia}}]{Zackrisson2016}
{Zackrisson}, E., {Binggeli}, C., {Finlator}, K., {et~al.} 2016, ArXiv
  e-prints, arXiv:1608.08217

\end{thebibliography}

\end{document}